\documentclass{aa}  

\usepackage{graphicx}
\usepackage{lscape}
\usepackage{amsmath}
\begin{document}

   \title{The HARPS search for southern extra-solar planets.\thanks{Based on observations taken with the HARPS spectrograph (ESO 3.6-m telescope at La Silla) under programmes 072.C-0488(E), 082.C-0212(B), 085.C-0063(A), 086.C-0284(A), and 190.C-0027(A). }\fnmsep\thanks{Radial velocity and stellar activity data are only available in electronic form at the CDS via anonymous ftp to cdsarc.u-strasbg.fr (130.79.128.5) or via http://cdsweb.u-strasbg.fr/cgi-bin/qcat?J/A+A/.}}

   \subtitle{XXXIX. HD175607, the most metal-poor G dwarf with an orbiting sub-Neptune}

   \author{A. Mortier\inst{1},
           J.P. Faria\inst{2,3},
           N.C. Santos\inst{2,3},
           V. Rajpaul\inst{4},
           P. Figueira\inst{2},
           I. Boisse\inst{5},
           A. Collier Cameron\inst{1},
           X. Dumusque\inst{6},
           G. Lo Curto\inst{7},
           C. Lovis\inst{8},
           M. Mayor\inst{8},
           C. Melo\inst{7},
           F. Pepe\inst{8},
           D. Queloz\inst{8,9},
           A. Santerne\inst{2},
           D. S\'egransan\inst{8},
           S.G. Sousa\inst{2},
           A. Sozzetti\inst{10},
        \and
           S. Udry\inst{8}
          }

   \institute{SUPA, School of Physics and Astronomy, University of St Andrews, St Andrews KY16 9SS, UK\\
        \email{am352@st-andrews.ac.uk}
        \and
        Instituto de Astrof\'{\i}sica e Ci\^encias do Espa\c co, Universidade do Porto, CAUP, Rua das Estrelas, 4150-762 Porto, Portugal\\
        \and
        Departamento de F\'{\i}sica e Astronomia, Faculdade de Ci\^encias, Universidade do Porto, Portugal\\
        \and
        Sub-department of Astrophysics, Department of Physics, University of Oxford, Oxford OX1 3RH, UK \\
        \and
        Aix Marseille Universit\'e, CNRS, LAM (Laboratoire d'Astrophysique de Marseille) UMR 7326, 13388, Marseille, France\\
        \and
        Harvard-Smithsonian Center for Astrophysics, 60 Garden Street, Cambridge, Massachusetts 02138, USA\\
        \and
        European Southern Observatory, Casilla 19001, Santiago, Chile\\
        \and
        Observatoire de Gen\`eve, Universit\'e de Gen\`eve, 51 ch. des Maillettes, CH-1290 Sauverny, Switzerland\\
        \and
        Institute of Astronomy, University of Cambridge, Madingley Road, Cambridge, CB3 0HA, UK\\
        \and
        INAF - Osservatorio Astrofisico di Torino, Via Osservatorio 20, I-10025 Pino Torinese, Italy
        }

   \date{Received July 6, 2015; Accepted November 2, 2015}


  \abstract
   {The presence of a small-mass planet (M$_p<$0.1\,M$_{Jup}$) seems, to date, not to depend on metallicity, however, theoretical simulations have shown that stars with subsolar metallicities may be favoured for harbouring smaller planets. A large, dedicated survey of metal-poor stars with the HARPS spectrograph has thus been carried out to search for Neptunes and super-Earths.}
   {In this paper, we present the analysis of \object{HD175607}, an old G6 star with metallicity [Fe/H] = -0.62. We gathered 119 radial velocity measurements in 110 nights over a time span of more than nine years.}
   {The radial velocities were analysed using Lomb-Scargle periodograms, a genetic algorithm, a Markov chain Monte Carlo analysis, and a Gaussian processes analysis. The spectra were also used to derive stellar properties. Several activity indicators were analysed to study the effect of stellar activity on the radial velocities.}
   {We find evidence for the presence of a small Neptune-mass planet (M$_{p}\sin i = 8.98\pm1.10$\,M$_{\oplus}$) orbiting this star with an orbital period $P = 29.01\pm0.02$\, days in a slightly eccentric orbit ($e=0.11\pm0.08$). The period of this Neptune is close to the estimated rotational period of the star. However, from a detailed analysis of the radial velocities together with the stellar activity, we conclude that the best explanation of the signal is indeed  the presence of a planetary companion rather than stellar related. An additional longer period signal ($P\sim 1400$\,d) is present in the data, for which more measurements are needed to constrain its nature and its properties.}
   {HD\,175607 is the most metal-poor FGK dwarf with a detected low-mass planet amongst the currently known planet hosts. This discovery may thus have important consequences for planet formation and evolution theories.}   \keywords{ planetary systems / stars: individual: HD175607 / techniques: radial velocities / stars: solar-type / stars: activity / stars: abundances
               }

   \authorrunning{Mortier, A. et al.}
   \maketitle

%

\section{Introduction}

Very early after the first exoplanets were discovered, it was suggested that stars with a higher metallicity have a higher probability of hosting a Jupiter-like planet than stars with lower metallicity \citep{Gon97}. This result was confirmed in a number of subsequent studies \citep[e.g.][]{San01,Fis05,John10,ME13}. Taken at face value, it favours planet formation theories based on the core-accretion model \citep[e.g.][]{Pol96,Mor09,Mor12}. According to this model, dust and grains coagulate to form planetesimals and combine to make larger cores and thus planets. Metal-rich stars and disks can form these cores more quickly, so they have time to accrete gas before the disk dissipates resulting in more gas giants around metal-rich stars.

For lower-mass planets (M$_p<$0.1\,M$_{Jup}$), such as Neptunes and (super-)Earths, the same correlation is not observed and the planet occurrence rate even appears to be independent of the host-star metallicity \citep[e.g.][]{Udry07,Sou11b,Buch15}. This is also in agreement with core-accretion theories; see, however, \citet{Adi12c} or \citet{Wang15}. Planet synthesis simulations based on the theories of core-accretion and planet migration showed that the correlation may even be reversed in the case of Earth-sized planets where stars with subsolar metallicities are favoured for harbouring an Earth-sized planet \citep{Mor12}.

For these reasons, a sample of 109 metal-poor stars was chosen for an extensive radial velocity survey with the HARPS spectrograph \citep{Mayor03} to search for Neptunes and (super-)Earths \citep{San14}. The targets in this survey are bright, chromospherically quiet FGK dwarfs with metallicities between $-2.0$ and $-0.4$\,dex. More details about this programme can be found in \citet{San14}.

To this date, no low-mass planets have been detected in this metal-poor sample, although there is a debate over one star,  \object{HD\,41248}, that shows clear signs of radial velocity variability. \citet{Jen13} reported on the existence of two planets orbiting this star, close to the 7:5 mean motion resonance. However, using the extended dataset coming from our large programme, these planets could not be confirmed \citep{San14}. One of the signals can clearly be seen in the activity indicators and is thought to be due to the stellar rotation and stellar spots on the surface of the star. The other signal could not be detected any more in an extended dataset and may have shown up as a result of the time sampling of the data or as a signature of differential rotation (though \citet{Jen14} reported that the signals are coherent over time).
 
This paper reports on the presence of at least one Neptune around one of the stars of the metal-poor HARPS survey, HD\,175607. In Sect. \ref{obs} we describe the observations made. Section \ref{star} presents the stellar properties. We analyse the stellar activity in Sect. \ref{act} and the radial velocities in Sect. \ref{RV}. We discuss our findings in Sect. \ref{Disc}.

\section{Observations}\label{obs}

\object{HD\,175607} was observed with the HARPS spectrograph on the 3.6-m telescope at La Silla Observatory. A total of 119 spectra over 110 nights were taken between July 2004 and October 2013 under different observing programmes\footnote{It was first part of a GTO run, then part of three smaller, metal-poor programmes and eventually part of the large programme.}. Most spectra were observed with an exposure time of 15 minutes. This is done to average out noise (signals) coming from short-term stellar oscillations\citep[e.g.][]{San04b}. When the large programme started in October 2012, if possible, we tried to obtain two spectra separated by several hours in one given night to reduce granulation effects, following the optimised observational strategies from \citet{Dum11}. Since the signals analysed in this work are on much longer timescales, we then averaged over these two measurements per night. The spectra have a mean signal-to-noise ratio of 104 around 6200\AA.

Radial velocities (RVs) were homogeneously derived using the HARPS Data Reduction Software (DRS). This pipeline cross-correlates the observed spectra with a mask representing a G8 dwarf (the spectral type of HD\,175607 is G6V). By fitting a Gaussian to the cross-correlation function (CCF), the value and uncertainty of the RV is determined \citep[e.g.][]{Bara96,Pepe02b}. We end up with 110 precise RV measurements with a mean error bar of 0.95 m\,s$^{-1}$, including photon, calibration, and instrumental noise. This mean error bar is slightly lower than the average error bar of all the stars in our sample. The data are taken over a time span of 3390 days (i.e. 9 years and 3 months).

\begin{figure}[t!]
\begin{center}
\includegraphics[width=\linewidth]{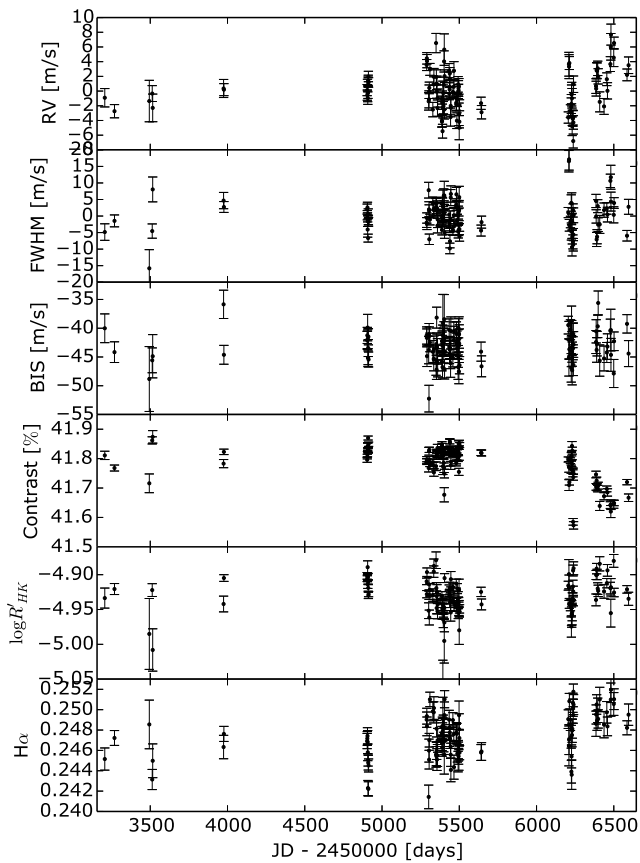}
\caption{Top to bottom: Time series of the radial velocities, FWHM, BIS, contrast, $\log{R'_{HK}}$, and H$\alpha$ index (the mean value is subtracted for the RVs and FWHM).}
\label{FigTime}
\end{center}
\end{figure}

From the DRS, we also get measurements for different stellar activity indicators: full width at half maximum (FWHM) of the CCF, line bisector inverse slope (BIS), contrast from the CCF, chromospheric activity indicator $\log{R'_{HK}}$ from the \ion{Ca}{ii} H\&K lines, H$\alpha$ index\footnote{The FWHM and contrast were corrected with a second-degree polynomial to account for the telescope losing focus over time}. Error bars for the FWHM, BIS, and contrast were scaled from the radial velocity error, following \citet{Sant15}. Figure \ref{FigTime} shows the radial velocity time series, together with the time series of all these indicators.

\section{Stellar properties}\label{star}

\begin{table}[t!]
\caption{Stellar parameters for HD175607.}
\label{TabSt}
\begin{tabular}{lcl}
\hline\hline
\noalign{\smallskip}
Parameter       & Value & Note \\
\hline
RA [h m s]      & 19 01 05.49 & (1) \\
DEC [d m s]     & -66 11 33.65 & (1) \\
Spectral type   & G6V & \\
$m_v$   & 8.61 & \\
$B-V$   & 0.70 & \\
Parallax [mas]  & 22.09$\pm$1.01 & (1) \\
Distance [pc]   & 45.27$\pm$2.07 & \\
$T_{\rm eff}$ [K]  & 5392$\pm$17 & (2) \\
$\log{g}$       & 4.64$\pm$0.03 & (2) \\
${\rm [Fe/H]}$  &$-$0.62$\pm$0.01 & (2) \\
${\rm [\alpha/Fe]}$ & 0.26 & (3) \\
Mass $[M_{\odot}]$ & 0.74$\pm$0.05 & (4) \\
Radius $[R_{\odot}]$ & 0.71$\pm$0.03 & (4) \\
Mass $[M_{\odot}]$ & 0.71$\pm$0.01 & (5) \\
Radius $[R_{\odot}]$ & 0.70$\pm$0.01 & (5) \\
Age [Gyr] & 10.32$\pm$1.58 & (5) \\
$<\log{R'_{\rm HK}}>$ & $-$4.92 & \\
P$_{Rot}$ [days] & 28.95$\pm$0.33 & (6) \\
P$_{Rot}$ [days] & 29.68$\pm$0.47 & (7) \\
$v\,\sin{i}$ [km~s$^{-1}$]      & 0.9 & (8) \\
$v\,\sin{i}$ [km~s$^{-1}$]      & 1.31 & (9) \\
\hline
\noalign{\smallskip}
\end{tabular}
\tablefoot{(1) \citet{VanL07}; (2) \citet{Sou11a}, with the surface gravity corrected following \citet{ME14}; (3) \citet{Adi12}; (4) using the \citet{Tor10} calibration; (5) Bayesian estimation \citep{Das06} using the PARSEC isochrones \citep{Bre12}; (6) using the empirical relationships from \citet[][their Eqs. 3 and 4]{Noy84b}; (7) using the empirical relationship from \citet[][their Eq. 5]{Mama08}; (8) \citet{Gleb05}; (9) using the recipe of \citet{San02}, adapted to the HARPS CCF }
\end{table}

HD\,175607 is a bright dwarf star of spectral type G6. It is located at a distance of $45.27$\,pc from the Sun, according to the new HIPPARCOS reduction \citep{VanL07}. All relevant stellar parameters can be found in Table \ref{TabSt}.

The stellar atmospheric parameters, effective temperature, surface gravity, and metallicity have been derived by a spectroscopic line analysis on a spectrum resulting from the sum of five individual HARPS spectra, with a total signal-to-noise ratio of 246.40 \citep{Sou11a}. Equivalent widths of iron lines (\ion{Fe}{I} and \ion{Fe}{II}) were automatically determined. These were then used, along with a grid of ATLAS plane-parallel model atmospheres \citep{Kur93}, to determine the atmospheric parameters, assuming local thermodynamic equilibrium in the MOOG code\footnote{\url{http://www.as.utexas.edu/~chris/moog.html}} \citep{Sne73}. More details on the method are found in \citet{Sou11a} and references therein.

They found a temperature of $5392\pm17$\,K. \citet{Cas11} used photometry to derive stellar parameters and obtained a slightly hotter temperature of 5521\,K. Given the known issues with the spectroscopic derivation of the surface gravity \citep[e.g.][]{Tor12,ME13c}, we corrected the surface gravity from \citet{Sou11a} to a more accurate value with the formula provided in \citet{ME14}. The spectroscopic metallicity of $-0.62 \pm 0.01$ shows that this star is indeed metal poor, although within the metal-poor survey, it belongs to the more metal-rich half of the sample. The presented errors are precision errors, intrinsic to the  spectroscopic method we used, and are very small. A discussion on the systematic errors of our method can be found in \citet{Sou11a}, their Sect. $3.1$. For effective temperature, a systematic error of $60$\,K is quoted while for metallicity, they quote a systematic error of $0.04$\,dex.

\citet{Adi12} calculated the chemical abundances of this star and found that it is alpha-enhanced (${\rm [\alpha/Fe]} = 0.26$). Kinematically this star would belong to the thin disk, or transitioning between the thin and thick disk \citep{Adi12}. The alpha-enhancement could hint that this star is more likely to be a planet host since \citet{Adi12b} found in the HARPS GTO and Kepler samples that iron-poor planet hosts (in all mass regimes) are alpha-enhanced, while single iron-poor stars show no enhancement in other metals.

Stellar masses and radii were derived using two methods. First, to maintain homogeneity with the online catalogue for stellar parameters of planet hosts \citep[SWEET-Cat\footnote{\url{https://www.astro.up.pt/resources/sweet-cat/}} -][]{San13}, we used the corrected calibration formulae of \citet{Tor10}\footnote{See \citet{San13} for details on the correction.}. This gives us a stellar mass of $0.74\pm0.05$\,M$_{\odot}$ and a stellar radius of $0.71\pm0.03$\,R$_{\odot}$. Second, we also used a Bayesian estimation of stellar parameters \citep{Das06} through their web interface\footnote{http://stev.oapd.inaf.it/cgi-bin/param}. For this, we used the apparent V magnitude, the Hipparcos parallax, the effective temperature and metallicity from the spectroscopic analysis, and the PARSEC isochrones \citep{Bre12}. From the models, we obtain a stellar mass  of $0.71\pm0.01$\,M$_{\odot}$ and a stellar radius of $0.70\pm 0.01$\,R$_{\odot}$ which are comparable with the results from the calibration formulae. Using the same input and through the same web interface for the Bayesian isochrone fitting \citep{Das06,Bre12}, we also get an estimate for the stellar age ($10.32$\,Gyr) that makes it a fairly old star. It also returns a value for the surface gravity, $4.57\pm0.01$, which is close to the corrected spectroscopic value. Since the isochronal stellar mass value is more precise, we  use that value for the duration of this paper.

HD\,175607 is a slowly rotating star. \citet{Gleb05} report a value for the projected rotational velocity $v\,\sin{i} = 0.9$\,km/s. Following a similar recipe in \citet{San02}, we used the B-V colour and the mean FWHM of all 110 measurements to obtain an estimate of $v\,\sin{i} = 1.31$\,km/s. We get an estimate for the rotational period with the empirical relationships of \citet[][their Eqs. 3 and 4]{Noy84b} or \citet[][their Eq. 5]{Mama08} via
the chromospheric activity indicator $\log{R'_{\rm HK}}$. The weighted mean value of $\log{R'_{\rm HK}}$ is $-4.92$ over all 110 measurements. Combining this with the B-V, we obtain an estimated rotational period of about 29 days. This is just an estimate resulting from calibrations and  the true rotational period is not known. All stellar parameters are in Table \ref{TabSt}.

\section{Activity analysis}\label{act}

Even in relatively inactive stars, radial velocity variations can be induced by stellar mechanisms other than orbiting planets, such as intrinsic stellar variations coming from stellar spots and/or faculae on the surface of the star \citep[e.g.][]{Boi11,Hay14,San14,Rob15}. It is thus important that we study the stellar activity  to be able to distinguish between RV signals coming from a planet and those from the star itself. As mentioned in Sect. \ref{obs}, we have measurements of different activity indicators. If periodic variations in the RV signal were also  present in one or more of these activity indicators, that could mean that the RV variation is activity induced rather than planet induced.

\begin{figure}[t!]
\begin{center}
\includegraphics[width=\linewidth]{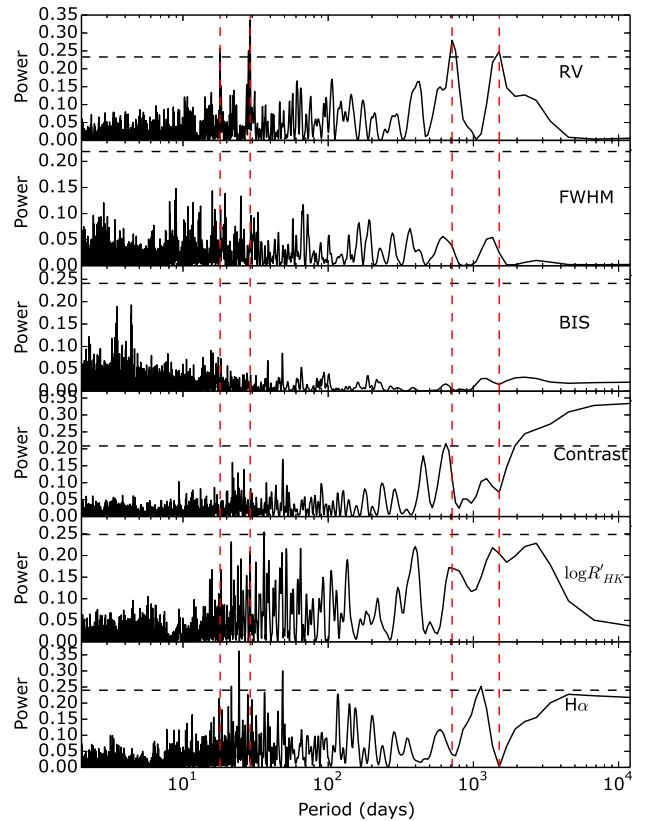}
\caption{Top to bottom: GLS periodograms of the radial velocities, FWHM, BIS, contrast, $\log{R'_{HK}}$, and H$\alpha$ index. The horizontal black dashed lines represent the $1\%$ FAP. The vertical red dashed lines appear at the periods of the four significant peaks in the RV periodogram.}
\label{FigPer}
\end{center}
\end{figure}

Figure \ref{FigPer} shows the General Lomb-Scargle (GLS) periodograms \citep{Zech09} from the RV and the four main activity indicators provided by the HARPS DRS pipeline: FWHM, BIS, contrast, and $\log{R'_{\rm HK}}$. A bootstrapping method is used to determine the $1\%$ false alarm probability \citep[FAP, for details see][]{Me12}. There are four significant peaks in the RV periodogram (see more in Section \ref{RV}). The most significant peak is seen around 29\,days, which is the same as the estimated rotational period from the activity level (see previous Section). Studying the activity indicators as proxies of stellar activity is thus even more important in this specific case.

When we look at the GLS periodograms of the CCF parameters (FWHM, BIS, contrast), none of the peaks seen in the RV periodogram are observed. In fact, none of these indicators show strong periodical patterns. There is some short-term (3-5 days), non-significant variation in the BIS, but none of these signals could be found in the RV periodogram. In fact, the estimated rotational period is not clear from these indicators. The periodogram of the H$\alpha$ index shows significant peaks at $24.5$ and $48$ days and some long-term variation. The significant periodicities from the RV periodogram cannot be seen here either.

Additionally, we computed other activity indicators, also derived directly from the CCF, using the code provided by \citet{Fig13}\footnote{'Line Profile Indicators': \url{http://www.astro.up.pt/exoearths/tools.html}}. We derived values for the BIS- and BIS+ \citep{Fig13},  Vspan \citep{Boi11}, and  biGauss \citep{Nar06}. All these indicators are used as alternatives to the BIS, but can probe the line profile variations better in case of low signal-to-noise ratio (e.g. BIS-, Vspan) or correlations close to the noise level (e.g. BIS+, biGauss). None of these indicators show significant variation or correlations with RV either.

By examining the patterns in the $\log{R'_{\rm HK}}$, we find that there is a forest of peaks in the GLS periodogram between 20 and 70 days, of which the peak around 36\,days is significant. However, the most significant peaks in the RV periodogram are not among the stronger peaks in the periodogram of $\log{R'_{\rm HK}}$. Furthermore, the same forest of peaks cannot be seen in the periodogram of the RVs. Additionally, there is some long-term variation in the $\log{R'_{\rm HK}}$ and contrast at periods that appear to be present in the RV data as well (see next section for further discussion on this).

\begin{figure}[t!]
\begin{center}
\includegraphics[width=\linewidth]{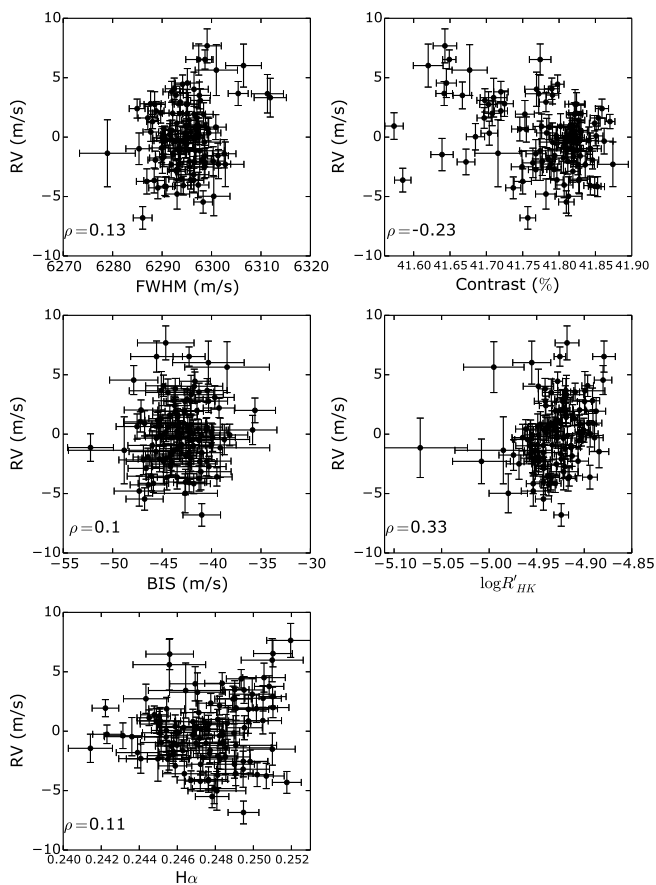}
\caption{Correlations between the RV (mean-subtracted) and the five main activity indicators: FWHM, contrast, BIS span, $\log{R'_{\rm HK}}$, and H$\alpha$. The Spearmann rank-order correlation coefficient is indicated in each panel. No significant correlations can be found.}
\label{FigAct}
\end{center}
\end{figure}

If the strongest variations in the RV were due to stellar activity, one can expect to find linear or figure-eight-shaped correlations between the RV and  activity indicators \citep[e.g.][]{Boi11,Fig13}, but the situation can also be more complex \citep{Dum14b}. Figure \ref{FigAct} plots the main activity indicators against the RV. No clear correlations can be seen among any of them. All (absolute) Spearman's rank correlation coefficients are lower than 0.3. The additional indicators we derived also showed no significant correlations. This makes us confident that the most significant peak in the RV is not due to activity and would be better explained by the presence of a planet. The fact that this peak is close to the estimated rotational period is discussed in Sect. \ref{Disc}.

\section{Radial velocity analysis}\label{RV}

\subsection{Periodograms}\label{Per}

In the previous section, we found that there are multiple significant periodicities in the RV data and that we have no reason to think that these are caused by stellar activity. As a first analysis, we performed a sequential pre-whitening on the RV data with GLS periodograms. We calculate the $1\%$ FAP level with a bootstrapping method. Then we identify the highest peak and the circular orbital solution creating that peak, as given by the periodogram analysis. We subtract this signal from the data and perform the same analysis on the residual data. We iterate this process until there are no significant peaks left in the periodogram of the residuals.

\begin{figure}[t!]
\begin{center}
\includegraphics[width=\linewidth]{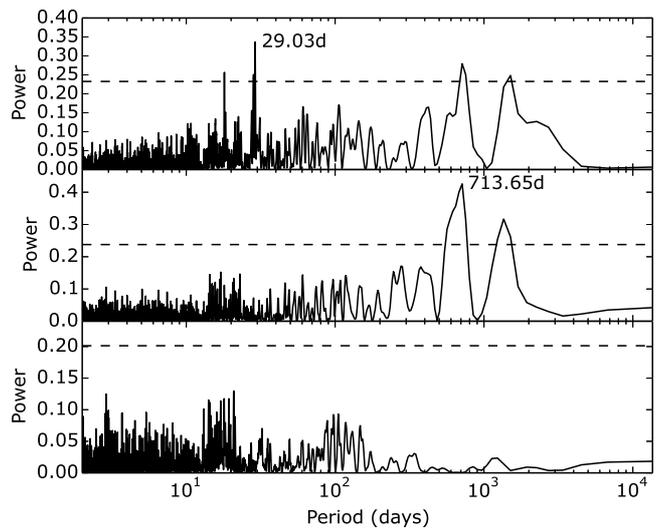}
\caption{Pre-whitening the radial velocities using GLS periodograms. Top panel: raw RVs. Middle panel: residual RVs after subtracting the best-fitted signal at $29.03$ days. Bottom panel: residual RVs after subtracting the best-fitted signals at $29.03$ and $713.65$ days. The horizontal black dashed lines represent the $1\%$ FAP.}
\label{FigPrew}
\end{center}
\end{figure}

Figure \ref{FigPrew} shows the results of this data pre-whitening. In the GLS periodogram of the original RV data, the strongest peak can be seen at $29.03$\,days. After removing this period from the data, we find that the peak at around 18\,days also disappeared. This hints at the fact that this period could be associated with the monthly alias of the 29-day period. The long-term periods are still significant, the highest of which is at $713.65$\,days. After subtracting this solution from the data, the other long-term period peak, at around 1400 days, also vanished. In the residual periodogram, the highest peak is now around 21\,days, but this is not significant and at the level of the noise. We thus find two significant periodicities in the data: one at 29 or 18 days and one at 713 or 1400 days.

\begin{figure}[t!]
\begin{center}
\includegraphics[width=\linewidth]{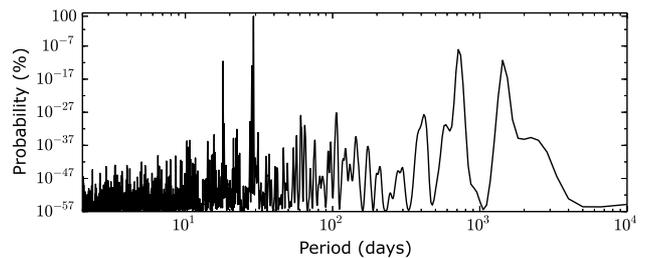}
\caption{BGLS periodogram of the raw RVs. The highest peak has been normalised to $100\%$ probability. This shows that the period at 29 days is $\sim 10^{10}$ times more probable than the period at 713 days.}
\label{FigBGLS}
\end{center}
\end{figure}

To assess the relative probability of the peaks in the periodograms, we used the Bayesian Generalized Lomb-Scargle Periodogram (BGLS) as described in \citet{ME15}\footnote{\url{https://www.astro.up.pt/exoearths/tools.html}}. Figure \ref{FigBGLS} shows this BGLS where the probability of the highest peak (at 29 days) is normalised at $100\%$. This analysis shows that the period at 29 days is $\sim 10^{10}$ times more probable than the period at 713 days. The periods have a median relative probability of $<P>\sim 10^{-55}\%$, so it is highly probably that  the observed periodicities are associated with real periodic signals in the data.

A multi-frequency periodogram \citep[e.g.][]{Balu13} can also be used to detect multiple periodicities in the data and assess their significance. We used FREDEC \citep[for details see][]{Balu13}. We looked for all tuples of significant periodicities in the data with periods between 2 and 10000 days. We find several significant possibilities for a two-period solution. The strongest solution, with a tuple FAP of $1.66\cdot10^{-7}\%$ (and the lowest $\chi^2$-value), is found for the combination of periods at 29 and 706 days. All combinations are made up of a short period (29 or 18 days) and a longer period (700 or 1400 days).

\subsection{Statistical analysis}

Periodograms are tools to check which sinusoidal periodicities are present in a dataset. They are important for a first interpretation of the data, but to get a more robust fit of the data and to assess error bars on the parameters, other methods should be employed. We used a genetic algorithm, an MCMC algorithm, and a Gaussian processes (GP) analysis.

\subsubsection{Genetic algorithm}

Initially, we ran a genetic algorithm using {\it yorbit} \citep{Seg11}. This algorithm uses a population of 4800 genomes where each genome (defined by frequency, phase, and eccentricity) corresponds to a planetary system. We ran the genetic algorithm twice, once assuming one planet and once assuming two planets. No conditions were set on any of the parameters. A restriction on the eccentricity is automatically set to avoid the planet colliding with the star. Initial starting positions are chosen based on the peaks in the periodogram. The evolution ended when more than $95\%$ of the population converged within 3 sigma of the best solution. 

The one planet model ended with a population of planets with periods $P=29.022 \pm 0.014$\,days and eccentricities $e=0.148\pm0.084$. For the two planet model, we again find this planet around 29 days ($P =29.007 \pm 0.014$ and $e=0.091 \pm 0.037$). The second planet, however, is not that well constrained. Similar to the frequency analysis carried out in Sect. \ref{Per}, the algorithm finds two types of solutions with periods equally distributed around 700 or 1400 days. The longer period would also be slightly more eccentric, but all solutions have an eccentricity lower than $0.6$.

\subsubsection{Markov chain Monte Carlo}

The solutions explored by the genetic algorithm do not provide a reliable statistical population from which to perform inference. It only provides a small parameter space that could be a good starting point for more robust fitting methods such as sampling from the posterior probability through MCMC. This alternative method allows the posterior distribution of each parameter to be inferred. We employ the following model for the RVs:

\begin{equation}
\mathrm{RV}(t) = \gamma + \sum_i K_i[\cos(\omega_i + \nu(t,e_i,T_{0,i},P_i)) + e_i \cos\omega_i],
\end{equation}where $\gamma$ is the constant systemic velocity, $K$ the RV amplitude, $e$ the eccentricity, $\omega$ the argument of periapse, and $\nu(t)$ the true anomaly. A sum is taken over all possible Keplerian signals. The true anomaly is a function of time, eccentricity, the period $P$, and the time of periastron passage $T_0$. It is defined as

\begin{equation}
\tan\frac{\nu}{2} = \sqrt{\frac{1+e}{1-e}}\tan\frac{E}{2}
,\end{equation}with $E$ the eccentric anomaly, which in turn can be found by solving Kepler's equation

\begin{equation}
E - e\sin E = 2\pi \frac{t-T_0}{P}.
\end{equation}

An additional jitter term is quadratically added to the error bars to incorporate the underestimation of these RV error bars and account for any additional noise present in the data. The final Gaussian likelihood function is

\begin{equation}
p(D|\boldsymbol\theta) = \prod_{i=1}^N \left[ \frac{1}{\sqrt{2\pi(\sigma_i^2 + \mathrm{jitter}^2)}}\exp\left(-\frac{[y_i - \mathrm{RV}(t_i)]^2}{\sigma_i^2 + \mathrm{jitter}^2}\right) \right],
\end{equation}where $N$ is the number of datapoints, $\boldsymbol\theta$ the set of parameters in the RV model, and $D$ the data. This data consists of the times of observation $t_i$, the measured radial velocities $y_i$, and the estimated error bars $\sigma_i$.

The parameter set $\boldsymbol\theta$ has a prior distribution $p(\boldsymbol\theta)$. We assume that all parameters are independent so that the total prior distribution can be expressed as the product of the prior distributions of each parameter. We take uniform priors for $\gamma$, $T_0$, $e$, and $\omega$, a Jeffreys prior for the period $P$, and a modified Jeffreys prior for the amplitude $K$ and the jitter term \citep[as in][]{Greg05}. The knee for this modified Jeffreys prior is taken to be the mean error bar $\bar{\sigma_i}$. All priors used for the MCMC are listed in Table \ref{tab:priors}.

\begin{table}[t!]
\caption{Priors for the MCMC procedure}
\label{tab:priors}
\begin{tabular}{llc}
\hline\hline
Parameter       & Prior & Limits \\
\hline
$\gamma$ [m/s]  & Uniform & [-91906.42, -91871.82] \\
jitter  & Mod. Jeffreys$^{\ast}$ & [0.0, 5.0] \\
$K_1$ [m/s]     & Mod. Jeffreys$^{\ast}$ & [0.0, 10.0] \\
$P_1$ [d]       & Jeffreys & [27.0, 32.0] \\
$e_1$   & Uniform & [0, 1[ \\
$\omega_1$      & Uniform &  \\
$T_0,1$ [JDB]   & Uniform & [2455500.0, 2455560.0] \\
$K_2$ [m/s]     & Mod. Jeffreys & [0.0, 10.0] \\
$P_2$ [d]       & Jeffreys & [200.0, 2000.0] \\
$e_2$   & Uniform & [0, 1] \\
$\omega_2$      & Uniform &  \\
$T_0,2$ [JDB]   & Uniform & [2454300.0, 2456300.0] \\
\hline
\end{tabular}
\tablefoot{$\ast$ Knee for the modified Jeffreys prior is taken to be the mean error bar $\bar{\sigma_i}$.}
\end{table}

Using Bayes' theorem, the posterior density is then expressed as

\begin{equation}
p(\boldsymbol\theta|D) = \frac{p(\boldsymbol\theta)p(D|\boldsymbol\theta)}{p(D)}.
\end{equation}Herein, the data probability $p(D)$ is seen as a normalisation constant and is kept at $1$ for the MCMC procedure. We calculate $p(D)$ later to compare the different models.

In the MCMC routine, we calculate the natural logarithm of the posterior probability density. Furthermore, we perform a coordinate transformation and use $\sqrt{e}\cos(\omega)$ and $\sqrt{e}\sin(\omega)$ instead of $e$ and $\omega$ \citep[see e.g.][]{Ford06}. This can be done easily since the Jacobian factor for this transformation is 1. To run the MCMC, we use {\it emcee} \citep{Fore13}, a Python code that implements an affine invariant MCMC ensemble sampler \citep{Good10}. An initial guess for the walkers is randomly chosen inside the final population of the genetic algorithm. We used 700 walkers  with 2000 steps. We allow for a burn-in period, which is chosen to be ten times the maximum autocorrelation time of the resulting walkers. Afterwards, we additionally perform a declustering method to remove the walkers with significantly lower posterior probabilities \citep[as in][]{Hou12}. This removes the walkers that got stuck inside local maxima.

\begin{figure}[t!]
\begin{center}
\includegraphics[width=\linewidth]{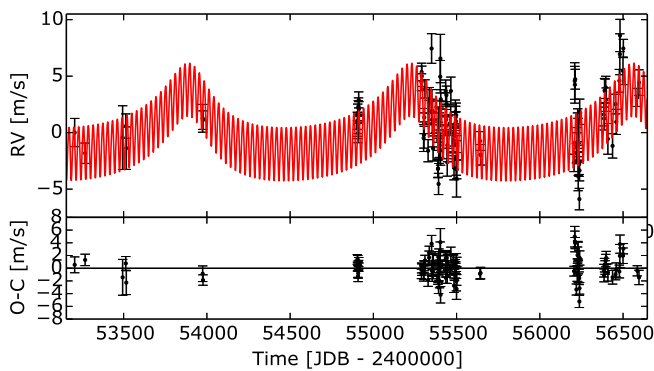}
\caption{Full orbit, using the MCMC results of a two Keplerian model. Top panel: relative RVs versus time; bottom panel: residuals.}
\label{Fig:orbitmcmc}
\end{center}
\end{figure}

\begin{figure}[t!]
\begin{center}
\includegraphics[width=\linewidth]{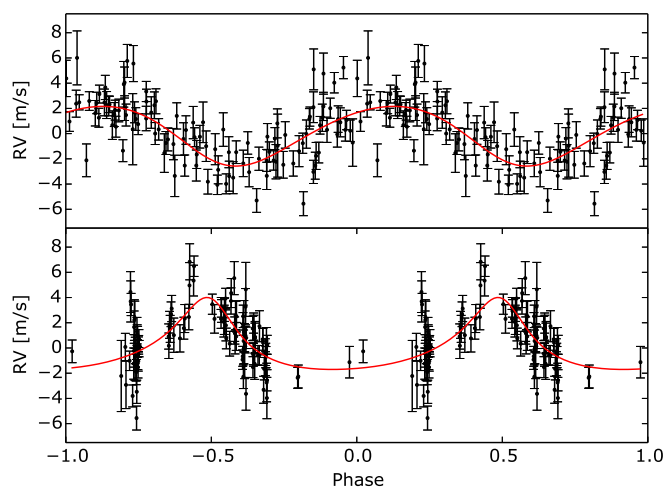}
\caption{Phased orbits, using the MCMC results of a two Keplerian model. Top panel: 29d signal; bottom panel: 1400d signal.}
\label{Fig:orbitphasemcmc}
\end{center}
\end{figure}

\begin{table*}
  \caption{Planetary parameters from the MCMC and GP fitting procedures. Errors are the $1\sigma$ uncertainties taken from the posterior distributions.}
  \label{tab:planet_fit}
  \centering
    \begin{tabular}{c|ccc|ccc|cc}
\hline\hline
Parameter	& \multicolumn{3}{c|}{MCMC - 1 planet}	& \multicolumn{3}{c|}{MCMC - 2 planets}	&  \multicolumn{2}{c}{GP - 1planet} \\
	& median	& $+\sigma$	& $-\sigma$	& median	& $+\sigma$	& $-\sigma$	& MAP value	& $\pm\sigma$ \\
\hline
$\gamma$ [m/s]	& $-91889.69$	& $0.22$	& $0.22$	& $-91890.41$	& $0.29$	& $0.28$	& $-91889$	& $1$ \\
& & & & & & & & \\
$K_1$ [m/s]	& $2.21$	& $0.33$	& $0.33$	& $2.37$	& $0.29$	& $0.30$	& $1.8$	& $0.4$ \\
$P_1$ [d]	& $29.03$	& $0.03$	& $0.03$	& $29.01$	& $0.02$	& $0.02$	& $29.0$	& $0.2$ \\
$m_1\sin i$ [M$_{\oplus}$] & $8.26$	& $1.25$	& $1.25$	& $8.98$	& $1.10$	& $1.10$	& $6.7$	& $1.5$ \\
$e_1$	& $0.16$	& $0.14$	& $0.11 $	& $0.11$	& $0.09$	& $0.07$	& $0.17$	& $0.10$ \\
$\omega_1$ 	& $0.55\pi$	& $0.36\pi$	& $0.34\pi$	& $0.79\pi$	& $0.29\pi$	& $0.29\pi$	& $1.0\pi$	& $0.40\pi$ \\
$T_0,1$ [BJD]	& $2455528.01$	& $4.71$	& $5.16$	& $2455532.17$	& $4.15$	& $4.18$	& $2453219$	& $6$ \\
& & & & & & & & \\
$K_2$ [m/s]	& --	& --	& --	& $2.86$	& $0.51$	& $0.51$	&  --	&  -- \\
$P_2$ [d]	& --	& --	& --	& $1336.61$	& $103.27$	& $45.50$	&  --	&  -- \\
$m_2\sin i$ [M$_{\oplus}$] & --	& --	& --	& $34.97$	& $6.93$	&	&  --	& -- \\
$e_2$	& --	& --	& --	& $0.42$	& $0.15$	& $0.14$	&  --	&  -- \\
$\omega_2$ 	& --	& --	& --	& $0.08\pi$	& $0.10\pi$	& $0.09\pi$	&  --	&  -- \\
$T_0,2$ [BJD]	& --	& --	& --	& $2455244.26$	& $63.39$	& $73.95$	&  --	&  -- \\
& & & & & & & & \\
jitter	& $2.01$	& $0.17$	& $0.19$	& $1.40$	& $0.16$	& $0.17$	& --	& --  \\
$P_{gp}$	& --	& --	& --	& --	& --	& --	& $29.9$	& $0.2$ \\
$\lambda_{\rm p}$ & --	& --	& --	& --	& --	& --	& $0.16$ 	& $0.02$ \\
$\tau$ [d] & --	& --	& --	& --	& --	& --	& $67$	& $11$ \\
\hline
\end{tabular}%


\end{table*}

Results for the one- and two-Keplerian models are listed in Table \ref{tab:planet_fit}. The best fit for the two Keplerian model is shown in Figs. \ref{Fig:orbitmcmc} and \ref{Fig:orbitphasemcmc}. A periodogram of the residuals reveals just noise, so it was chosen not to run a model with three Keplerians.

In order to compare the two models statistically, one would want to assess the Bayes factor, i.e.~the ratio of the model evidence. In the case of two models $M_1$ and $M_2$, each with the parameter set $\boldsymbol\theta_1$ and $\boldsymbol\theta_2$, the Bayes factor to assess model two over model one is expressed as:

\begin{equation}
B_{21} = \frac{P(D|M_2)}{P(D|M_1)} = \frac{\int P(\boldsymbol\theta_2|M_2) P(D|\boldsymbol\theta_2,M_2) d\boldsymbol\theta_2}{\int P(\boldsymbol\theta_1|M_1) P(D|\boldsymbol\theta_1,M_1) d\boldsymbol\theta_1}.
\end{equation}

Calculating these integrals over the complete parameter space is tricky. However, there are ways to solve it. The {\it emcee} package provides a parallel-tempering ensemble sampler that can be used to estimate this integral. It makes use of thermodynamic integration as described in \citet{Gog04}. For a more detailed calculation, see Appendix \ref{App:PT}. We applied this formalism, using 20 different temperatures (each one increasing with $\sqrt{2}$) with 200 walkers each. As a burn-in, we used 1000 steps and then an additional 2000 steps for the integral calculation. We find that $B_{21}\sim \exp(15)$, supporting the model with two Keplerians with very strong evidence \citep[e.g.][]{Kass95}.

We emphasize that this evidence is dependent on the chosen priors. Specifically, the prior on the period of the inner planet may be seen as too narrow. We ran tests where the prior on this period is 1 to 100 days. We get comparable results as with the more narrow prior,  although the time of periastron (whose prior is then also widened) is less constrained because it is cyclic. Thermodynamic integration with these wider priors gives us a Bayes factor $B_{21}\sim \exp(19)$, even higher than before. We can thus be confident that the strong evidence is not due to our choice of priors.

\subsubsection{Gaussian processes}

Gaussian processes provide a mathematically-tractable and flexible framework for performing Bayesian inference about functions. They are particularly suitable for the joint modelling of deterministic processes (such as signals induced by planets) with stochastic processes of unknown functional forms such as activity signals \citep{Aig12, Hay14}. Despite not knowing the functional form of these stochastic processes, we usually know some of its properties.

\citet{Raj15}, hereafter R15, developed this kind of framework to  model  RV time series jointly with one or more ancillary activity indicators. This allows the activity component of the RV time series to be constrained and disentangled from planetary components. Their framework treats the underlying stochastic process, giving rise to activity signals in all available observables (RVs and ancillary time series) as being described by a GP, with a suitably-chosen covariance function. They then use physically-motivated and empirical models to link this GP to the observables; with the addition of noise and deterministic components (e.g.\ dynamical effects for the RVs), all observables can be modelled jointly as GPs, with the ancillary time series thus serving to constrain the activity component of the RVs. They showed their framework can be used to disentangle activity and planetary signals. This is the found even when the planetary signal is much weaker than the activity signal ($\Delta {\rm RV}\lesssim0.5$~m/s) and has a period {identical} to the activity signal. Since the period of the first signal in the data for \object{HD175607} is very close to the estimated rotational period of the star, we performed a fit for this signal with the GP framework as described in R15.

The marginal likelihood $\mathcal{L}(\boldsymbol\theta,\boldsymbol\phi)$ for the data, given a GP model, can be expressed as 

\begin{equation}
\ln \left[ {{\mathcal{L}}(\boldsymbol\theta,\boldsymbol\phi)} \right] =  - \frac{1}{2}{{\mathbf{r}}^{\text{T}}}{{\mathbf{K}}^{ - 1}}{\mathbf{r}} - \frac{1}{2}\ln \left( {\det {\mathbf{K}}} \right) - \frac{N}{2}\ln \left( {2\pi } \right),
\end{equation}where $\mathbf{r}(\mathbf{t},\boldsymbol\theta)= {\mathbf{y}} - {\mathbf{m}(\mathbf{t},\boldsymbol\theta)}$ is the vector of residuals of the data after the mean function $\mathbf{m}$ has been subtracted and $N$ is the number of datapoints. The free {hyper-parameters} $\boldsymbol\theta$ and $\boldsymbol\phi$ can then be varied to maximise $\mathcal{L}$; this process is known as Type-II maximum likelihood, or marginal likelihood maximisation \citep{Gibs12}. In so doing, we refine vague distributions over many, very different functions, the forms of which are controlled by $\boldsymbol\theta$ and $\boldsymbol\phi$, to more precise distributions that are focused on functions that best explain our observed data. 

We implemented the GP framework exactly as described in R15. In particular, given that we have a physical reason to expect a degree of periodicity in the activity signals (as they are modulated by the periodic rotation of the star), we adopted the following quasi-periodic covariance function for the framework's underlying, activity-driving process. This covariance function was previously considered by \citet{Aig12} to model observed variations in the Sun's total irradiance, and by \citet{Hay14} to model correlated noise in the \object{CoRoT-7} data

\begin{equation}
\label{eq:cov_qp}
 {k(t,t')} \propto \exp \left\{ - \frac{\sin^2 \left[\pi (t-t')/P_{gp}\right]}{2 \lambda_{\rm p}^2} 
 - \frac{(t-t')^2}{2\tau^2} \right\},
\end{equation}where $P_{gp}$ and $\lambda_{\rm p}$ correspond to the period and length scale of the periodic component of the variations and $\tau$ is an evolutionary timescale. While $\tau$ has units of time, $\lambda_{\rm p}$ is dimensionless. 

For HD176507, we jointly modelled  the $\Delta {\rm RV}$ (after subtracting a polynomial to exclude longer period variations), $\log R'_{\rm HK}$~ and BIS~time series as in R15. We chose not to include the FWHM since FWHM data are noisier than, but often very tightly correlated with $\log R'_{\rm HK}$, and thus often do not contain useful extra information that the other indicators have not yet provided.

Non-informative priors (just as for the MCMC procedure) were placed on all Keplerian orbital parameters (incorporated into the GP's mean function, $m$). These priors and the priors on the hyper-parameters are listed in Table \ref{tab:priorsGP}. Parameters for the Keplerian orbit are estimated using the MultiNest nested-sampling algorithm \citep{Feroz08,Feroz09,Feroz13}, with the GP hyper-parameters first fixed at their MAP values, as per the computational approximation motivated in \citet{Gibs12}. 

\begin{table}[t!]
\caption{Priors for the GP procedure}
\label{tab:priorsGP}
\begin{tabular}{llc}
\hline\hline
Parameter       & Prior & Limits \\
\hline
$\gamma$ [m/s]  & Uniform & [-91906.42, -91871.82] \\
$K$ [m/s]       & Mod. Jeffreys$^{\ast}$ & [0.0, 10.0] \\
$P$ [d] & Jeffreys & [27.0, 32.0] \\
$e$     & Uniform & [0, 1[ \\
$\omega_1$      & Uniform & [0, $2\pi$] \\
$T_0$ [JDB]     & Uniform & [$2453206.0, 2453206.0+P$] \\
$P_{gp}$        & Uniform & [1, 100] \\
$\lambda_{\rm p}$ & Jeffreys & [0.01, 100] \\
$\tau$ & Jeffreys & [0.1, 1000] \\
\hline
\end{tabular}
\tablefoot{$\ast$ Knee for the modified Jeffreys prior is taken to be the mean error bar $\bar{\sigma_i}$.}
\end{table}

\begin{figure*}
\sidecaption
\includegraphics[width=12cm]{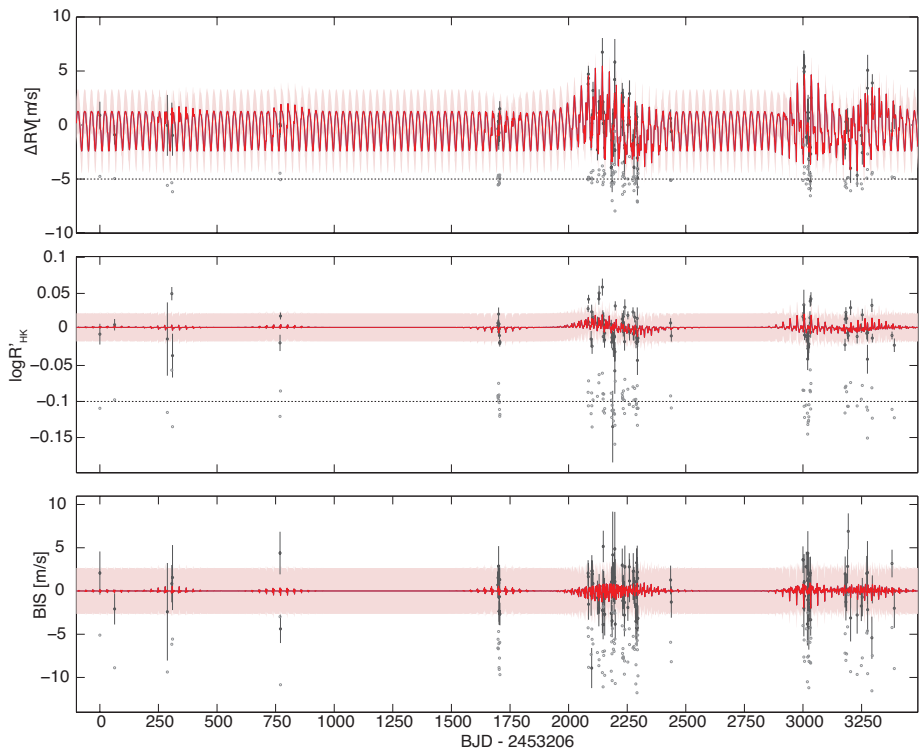}
\caption[]{GP model MAP fit to the  HD~175607 data. The $110$ observations in each time series were fit simultaneously, i.e.\ using a single set of (hyper)parameters. The dots indicate observed data, with estimated errors;  solid lines are model posterior means; and  shaded regions denote $\pm\sigma$ posterior uncertainty. Residuals are plotted below the observed data and fitted model, but for the sake of clarity, with an arbitrary vertical offset from the main time series.}
\label{fig:GP_fit_overall}
\end{figure*}

\begin{figure*}
\sidecaption
\includegraphics[width=12cm]{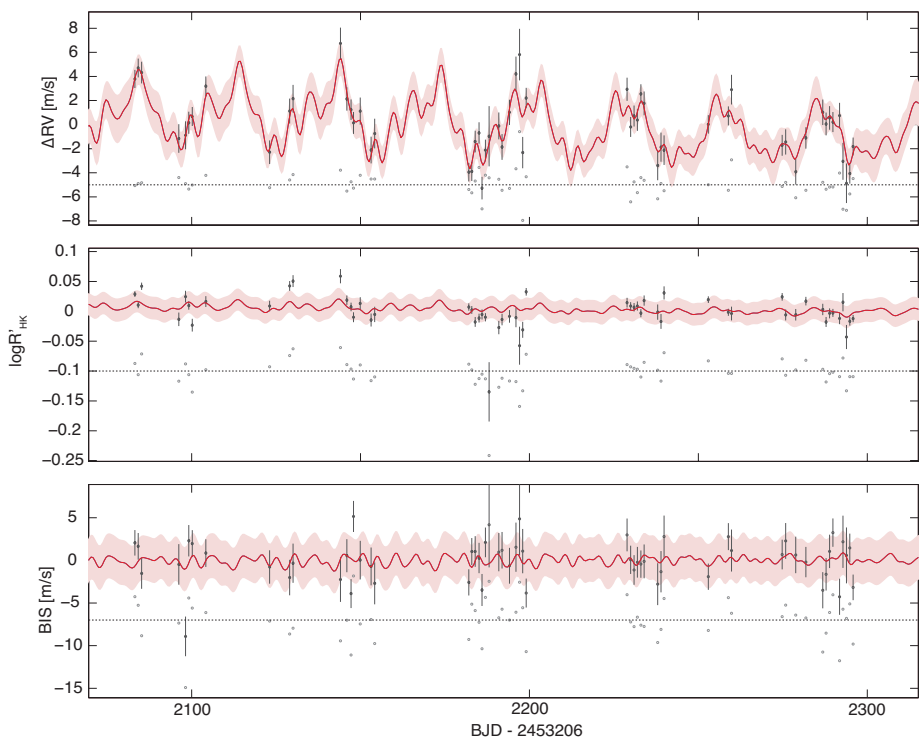}
\caption[]{GP model MAP fit to the  HD~175607 data, with a close-up view of the region of densest time coverage (57 observations over the course of about seven months). The dots indicate observed data, with estimated errors;  solid lines are model posterior means; and the shaded regions denote $\pm\sigma$ posterior uncertainty. Residuals are plotted below the observed data and fitted model, but for the sake of clarity, with an arbitrary vertical offset from the main time series.}
\label{fig:GP_fit_closeup}
\end{figure*}

Our findings were as follows. When not including a planetary component in the GP's mean function for the $\Delta {\rm RV}$~time series, the MAP value of the hyper-parameter $P_{gp}$ ended up being $29.0\pm0.1$~d: because the $29.0$-d signal was so significant in the $\Delta {\rm RV}$~time series, the GP was forced to absorb this, whilst all but ignoring and thus failing to fit the ancillary time series.

On the other hand, when including a Keplerian component, the hyper-parameter $P_{gp}$ ended up being $29.9\pm0.4$~d, with the $29$-d signal being absorbed entirely by the Keplerian component; under this model, the rms of the RV variations absorbed by the GP was reduced to the order of tens of centimetres per second. This is significant because whereas a GP can in principle model an arbitrarily-complex signal arbitrarily well  (the key constraint in R15's framework, however, is that the same quasi-periodic GP basis functions must be used to model RV and ancillary time series simultaneously), a Keplerian function is far simpler, and is always be strictly periodic. Therefore, the fact that the simpler, less flexible Keplerian interpretation is favoured by the GP framework indicates that the $29$~d signal must have a coherent phase over the entire dataset, strengthening the planetary interpretation of the $29$-d signal. The planet parameters we inferred when using the GP framework are presented in Table \ref{tab:planet_fit}. The evolution timescale for the activity signal is found to be $67$~d, slightly more than two rotation periods, as would be expected for this type of star.

We used the sample size-adjusted Akaike information criterion \citep[AIC;][]{Burn02} to select between the one-planet vs.\ no-planet models. The AICc value for the no-planet model was $-25.44$, and the corresponding value for the one-planet model $-33.06$, indicating that the planetary explanation was favoured by about a factor of ten.

The MAP fit using the one-planet model is presented in Fig.\ \ref{fig:GP_fit_overall} with a close-up in Fig.\ \ref{fig:GP_fit_closeup}. 
After subtracting the one-planet GP model, the residual~time series appeared white and normally-distributed, with no significant power on timescales smaller than one year, and with rms 0.95 m/s. This suggests that all of the RV variation (at least on timescales smaller than one year) can be explained fully with the planet + activity model. The $\log R'_{\rm HK}$ and BIS residuals contained no significant periodicities on any timescales.

\section{Discussion and conclusion}\label{Disc}

In this work we analysed the radial velocities of HD175607, a metal-poor ([Fe/H]=-0.62) dwarf star. These radial velocities show a clear periodicity around 29 days and a significant longer period signal. The main question is whether these signals are caused by a planet or rather another phenomenon resulting from the star itself. We  discuss each signal below.

\subsection{Short period signal}

The short period signal arises at 29 days. However, the rotational period is also estimated to be around 29 days and the Moon's orbital period is also close to 29 days, so caution is recommended. If this is due to a planet that would make the planet a small Neptune (M$_{p}\sin i = 8.98\pm1.10$\,M$_{\oplus}$ if the two-planet model is assumed).

Radial velocities can be contaminated by scattered light from the Moon. Specifically, this contamination can produce an additional dip in the CCF. If the Moon's velocity is close to the mean stellar velocity, the two dips are blended, which  affects the RV measurement of the star. In the case of \object{HD\,175607}, the mean stellar velocity is about $-92$\,km/s. The Moon orbits the Earth at about $1$\,km/s, and the Earth orbits the Sun at about $30$\,km/s. Consequently, the additional CCF dip due to scattered moonlight contamination is always going to be equal or more than $60$\,km/s redwards of the stellar CCF. This makes moonlight contamination in the RVs of this star impossible.

We emphasize that even if there would be contamination from the moon in our RVs, \citet{Cun13} showed that for the spectral type and magnitude of HD\,175607, the contamination would be around $10$\,cm/s, which is much lower than the signal seen here. We are thus confident that the $29$\,day signal is not due to the Moon.

Then remains the question of the rotational period. For several reasons listed below, we think that the signal is indeed best explained as being from a planet rather than activity-related:

\begin{itemize}
\item No significant correlations are found with any of the activity indicators provided by the HARPS DRS pipeline, nor with the extra activity indicators we calculated using the code in \citet{Fig13}. If the signal were to be activity related, one would expect there to be some correlation with at least one of the activity indicators. The lack thereof suggests the signal is planet related. 
\item The H$\alpha$ index shows significant periodicities around 24 days. It could thus be that the estimated rotational period, coming from the B-V colour and the mean $\log{R'_{\rm HK}}$, is not accurate and the rotational period is closer to 24 days. In this case, the RV signal would not be at the same period of the stellar rotation.
\item We have data spanning over nine years with about 4.5 years of intense datasets. This is of the order of 50 times the lifespan of a typical solar active region. Signals arising from activity are not expected to stay stable over this amount of time for this type of star. Since the period of the signal is still very well constrained, that  hints that the signal is stable over time and thus not due to activity.
\item In the GP analysis, the red noise is modelled separately from the Keplerian, though both are at similar periods. This analysis thus prefers the presence of a planet despite activity signals at similar periodicities. The planetary mass is lowest when using this model. We think this is because some of the signal's amplitude, swallowed by the GP, is treated as planetary in the other models.
\item If a signal is not stable over time, such as one caused by activity, the peak in a periodogram would be variable, depending on the amount of activity on certain times. We tested this and the peak gets always stronger when adding more data.
\item As a final test, we wanted to know what the expected periodogram power would be if we inject a noiseless Keplerian signal in the data with similar period, semi-amplitude, and eccentricity as the current signal. We thus injected a sinusoid with the same semi-amplitude and eccentricity but at a period of 21 days. As expected, we see a peak at 21 days. It has about the same power as the 29d peak. Since we did not add any noise for the 21d signal, this again hints that the 29\,d signal is of planetary nature.
\end{itemize}

There are other known cases where the orbital period is the same as the stellar rotation period, such as \object{CoRoT-11}b \citep{Gan10} or \object{XO-3}b \citep{Hebr08}. However, these are all cases of close-in hot Jupiters around fast-rotating stars, where the synchronous planetary orbit may come from tidal locking with the host star \citep[e.g.][]{Lan11, Bol12}. The 29d period of our mini-Neptune makes it implausible that the planet would have synchronised its host star since timescales for such a synchronisation scale with $(a/R_{\ast})^5\cdot1/M_p$ \citep[e.g.][]{Dob04,Brow11}. The planet could be tidally locked to the star, but there is no way of verifying that without the planetary spin period. There are several discovered planets with periods between 10 and 40 days, which are the typical orbital periods for slowly rotating stars, making it not that unlikely that some of them  have periods close to or similar to their estimated stellar rotational periods.

\subsection{Long period signal}

The long period signal is not as well constrained as the shorter period signal. From the MCMC, it was clear that the likelihood of a 1400d signal was much higher than the one from a 700d signal. The latter periods were sampled by the MCMC, but eventually removed in the declustering due to too low likelihood. If due to a planet, this planet would have a period of 1337 days and a minimum mass of about 35 Earth masses (i.e. 0.1 Jupiter masses), making it a large Neptune.

Though not statistically significant, similar long periodicities can be seen in the $\log R'_{\rm HK}$ and contrast. However, after removing the inner planet, there is still no significant correlation between the residual RVs and these indicators, nor did it get stronger. If the longer period signal were due to activity, we would have expected the correlations to arise when removing the shorter period signal.

With the long data span, we cover about 2.5 orbits of $\sim$1400 days. However, given the small number of datapoints in the first half of the dataset, we actually only span one full orbit. Furthermore, there are large gaps without data. We would need more data in order to confirm the nature of this signal and better constrain it in case of a planet. Follow up measurements are planned to resolve this.

\subsection{Metal-poor survey}

This detection is part of a large survey with the HARPS spectrograph for Neptunes around metal-poor FGK dwarfs. \object{HD\,175607b} is the first Neptune-mass planet discovered in this survey. Despite the low metallicity of the host star([Fe/H] = $-0.62$), it still belongs to the more metal-rich part of the sample. The metallicities for the entire sample range from $-1.5$ to $-0.05$\,dex \citep{San14}. In a forthcoming paper (Faria et al., submitted), the stars from this sample with more than 75 measurements, including HD\,175607, are discussed. Neptune-mass planets with periods lower than 50 days can be ruled out for these stars. 

In the literature, there are only few examples of Neptunes or super-Earths orbiting such metal-poor stars. The planetary system around \object{GJ\,667C} is one of them. It contains several super-Earths, while the star has a measured metallicity of $-0.55$\,dex \citep{Delf13,Rob14}. This star is an M-dwarf however and thus much cooler than HD\,175607. Another Neptune system is claimed around \object{Kapteyn}'s star \citep{Angl14,Bon13,Rob15b}, a very metal-poor ([Fe/H] = $-0.86$) halo star. This star is also an M-dwarf.

In this sense, HD\,175607 would be the most metal-poor FGK dwarf to date with an orbiting Neptune. Giant planets are also rare around metal-poor stars and it has been proposed that a lower metallicity limit ($\sim -0.7$) could exist for the formation of giant planets \citep{Me12}. Could the same be true for Neptunes or are we just still limited in the detection of lower-mass planets? This discovery may thus have important consequences for planet formation and evolution theories.

\begin{acknowledgements}

      This work made use of the Simbad Database. The research leading to these results received funding from the European Union Seventh Framework Programme (FP7/2007-2013) under grant agreement number 313014 (ETAEARTH). JPF acknowledges support from FCT through grant reference SFRH/BD/93848/2013. This work was supported by Funda\c{c}\~ao para a Ci\^encia e a Tecnologia (FCT) through the research grant UID/FIS/04434/2013. P.F., N.C.S., and S.G.S. also acknowledge the support from FCT through Investigador FCT contracts of reference IF/01037/2013, IF/00169/2012, and IF/00028/2014, respectively, and POPH/FSE (EC) by FEDER funding through the programme ``Programa Operacional de Factores de Competitividade - COMPETE''. This work results within the collaboration of the COST Action TD 1308. A.S. is supported by the European Union under a Marie Curie Intra-European Fellowship for Career Development with reference FP7-PEOPLE-2013-IEF, number 627202.

\end{acknowledgements}

\bibliographystyle{aa} 
\bibliography{/home/annelies/My_Articles/References.bib}

\begin{appendix}

\section{Model evidence from thermodynamic integration}\label{App:PT}

In this section, we describe how an estimate of the model evidence can be determined using thermodynamic integration. We define the temperature-evidence function $E(\beta)$ as

\begin{equation}\label{eq-ebeta}
E(\beta) = \int L^{\beta}(x)pr(x)dx,
\end{equation}where $L(x)$ is the likelihood, $pr(x)$ the prior, and $\beta = 1/T$ with $T$ the temperature. The model evidence that we want to compute is equal to $E(1)$. Also, $E(0)$ is equal to the integrated prior. Since normalised priors are used in this work, this integrated prior $E(0)$ is equal to $1$.

By using the formula for the differentiation of a natural logarithm,

\begin{equation}
\frac{d\ln E}{d\beta} = \frac{1}{E(\beta)} \frac{dE(\beta)}{d\beta},
\end{equation}

and plugging in Equation \ref{eq-ebeta}, we can write

\begin{equation}
\frac{d\ln E}{d\beta} = \frac{1}{E(\beta)} \int \ln L(x)L^{\beta}(x)pr(x)dx.
\end{equation}

The right-hand side of this equation is the average of the natural logarithm of the likelihood over the posterior at temperature $T = 1/\beta$. This is expressed as $\langle \ln L \rangle_{\beta}$

\begin{equation}
d\ln E = \langle \ln L \rangle_{\beta} d\beta.
\end{equation}

If we now integrate both sides of this equation over the interval $[0,1]$, we get 

\begin{equation}
\ln E(1) = \int_0^1 d\ln E = \int_0^1 \langle \ln L \rangle_{\beta} d\beta.
\end{equation}

This integral can be estimated from the parallel-tempering ensemble sampler, embedded in {\it emcee}. For each temperature, the average logarithm of the likelihood is estimated from the chains. The integral can then be estimated  using these values and applying a quadrature formula. From the estimation of the integral, we finally estimate the model evidence $E(1)$.

\end{appendix}

\end{document}